# Precipitable Water Vapor, Temperature, and Wind Statistics At Sites Suitable for mm and Submm Wavelength Astronomy in Northern Chile

Otarola, A.[1], De Breuck, C.[2], Travouillon[1,7], T., Matsushita S.[3], Nyman, L-Å.[4], Wootten, A.[5], Radford, S.J.E.[6], Sarazin, M.[2], Kerber, F.[2], Pérez-Beaupuits, J.P.[4]

[1] Thirty-Meter International Observatory, 100 W Walnut Street, Pasadena, CA, 91124, USA

[2] European Southern Observatory, Karl-Schwarzschild-Straße 2, 85748 Garching bei München, Germany

[3] Academia Sinica, Institute of Astronomy and Astrophysics, 11F, Astro-Math Bldg, AS/NTU, No.1, Sec. 4, Roosevelt Road., Taipei 10617, Taiwan, R.O.C.

[4] European Southern Observatory, Alonso de Córdova 3107, Vitacura, Santiago, Chile

[5] National Radio Astronomy Observatory, 520 Edgemont Road, Charlottesville, VA, 22903, USA

[6] Smithsonian Astrophysical Observatory, Submillimeter Array, 645 A'ohoku Place, Hilo, HI 96720 USA

[7] Research School of Astronomy and Astrophysics, Australian National University, Canberra, ACT 2611, Australia

*Keywords*: atmospheric effects, precipitable water vapor, site testing – submillimeter: general

Corresponding author: Angel Otarola, aotarola@tmt.org

*Abstract*

Atmospheric water vapor is the main limiting factor of atmospheric transparency in the mm and submm wavelength spectral windows. Thus, dry sites are needed for the installation and successful operation of radio astronomy observatories exploiting those spectral windows. Temperature and wind are variables of special consideration when planning the installation and operation of large-aperture radio telescopes, as these parameters affect the mechanical response of radio telescopes exposed to the environmental conditions. Temperature, and in particular temperature gradients, induce thermal deformation of mechanical structures, while high wind speeds and gusts induce pointing jitter affecting the tracking accuracy of astronomical sources during the observations. This work summarizes the statistics of precipitable water vapor (PWV), temperature, and wind monitored at sites by the coastal mountain range, as well as on the west slope of the Andes mountain range in the region of Antofagasta, Chile. This information could prove useful for the planning of extended baselines for the Atacama Large Millimeter and

Submillimeter Array (ALMA), and/or new radio telescope projects, such as the Atacama Large Aperture Submm/ mm Telescope (AtLAST) initiative.

## 1. Introduction

The Atacama desert in the northern part of Chile hosts some of the most advanced astronomical facilities, such as the European Southern Observatory (ESO) Very Large Telescope (VLT; Querel, R., & Kerber, F., 2014) located at the summit of Cerro Paranal, and the Atacama Large Millimeter/Submillimeter Array (ALMA; Wootten and Thompson, 2009) deployed at the Chajnantor plateau on the west slope of the Andes mountain range. Jointly, VLT and ALMA cover numerous spectral bands of astronomical interest in the visible, near-infrared, mid-infrared as well as submillimeter (submm) and millimeter (mm) regions of the electromagnetic spectrum. Near Paranal and Chajnantor, there are also sites hosting various other telescopes dedicated to large surveys of the sky in the near-infrared and observations of the microwave cosmic background radiation. Besides, ESO has already started the construction of a large-aperture telescope, the European Extremely Large Telescope (ELT) with an aperture of 39 m, intended for imaging and spectroscopy in the visible to mid-infrared, and equipped with adaptive optics instrumentation to enable diffraction limited capabilities in the near- and mid-infrared spectral bands. Yet, there are important scientific cases that would benefit from a larger field of view, high angular resolution, as well as high sensitivity for imaging and spectroscopy research in the millimeter and submillimeter spectral bands. In particular, there are scientific goals that can only be accomplished by, or highly benefit from, observations using a larger diameter[1] single-dish radio telescope (see, for instance, L. Testi[2] 2015; Kawabe et al., 2016). With these goals in mind, over 100 members of the international radio astronomy community, organized under various working groups, met at a workshop organized at the ESO headquarters to share ideas and priorities on the specific scientific cases that would highly benefit from the availability of a large-aperture radio telescope operating in the mm and submm spectral bands. This workshop was organized as part of the Atacama Large-Aperture Submm/ mm Telescope (AtLAST) initiative[3].

This work, prepared by the AtLAST initiative Site Working Group, presents the results from the analysis of existing site testing and telescope operations weather data. These results may help with the identification of suitable sites for the deployment of a large-aperture telescope intended for observations in the mm and submm spectral bands, in particular a telescope as that intended in the AtLAST initiative. However, the information is relevant to any group in search of a site for the deployment of new astronomical facilities that may consider the Atacama Desert region as an option.

---

[1] As for a large diameter mirror telescope or large diameter dish radio telescope we use also the term "large aperture" telescope or radio telescope, respectively.

[2] Testi, L., et al., 2015, ESO Submm Single Dish Scientific Strategy WG Report, available at https://www.eso.org/public/about-eso/committees/stc/stc-87th/public/STC-567_ESO_Submm_Single_Dish_Scientific_Strategy_WG_Report_87th_STC_Mtg.pdf

[3] https://www.eso.org/sci/meetings/2018/AtLAST2018.html

The sites in this geographical region, for which weather data was made available to this study, are listed in Table 1 and described in section 2. Atmospheric parameters, such as atmospheric precipitable water vapor, are important for the determination of the atmospheric transmission in various spectral bands of interest, while wind and temperature are relevant for considerations in the telescope design as well as in telescope operations. The analysis of these weather parameters, data, and results, are covered in Section 3. The main conclusions and recommendations are summarized in Section 4.

## 2. Sites: selected sites of great potential for astronomy projects in the Region of Antofagasta (Chile)

The dryness of the atmosphere over the Atacama region is explained by various factors, the chief ones are as follows. a) Its tropical latitude corresponding to the southern subsiding branch of the Hadley cell circulation. The descendent air is dry because most of its moisture condensed out in the earlier sections of the Hadley cell circulation, closer to the equatorial latitudes. This helps maintain in the area a semi-permanent anticyclone that is responsible of heating the lower and mid atmosphere. b) Its westerly location with respect to the high Andes mountain range that protects the area from the advection of humid air from the Atlantic basin. c) The existence of a cold ocean current flow along the east coast of the Pacific that helps create a temperature inversion (Muñoz et al., 2011) that keeps the relatively humid air at altitudes below the peaks of the coastal mountain range. Evidence of this sharp temperature inversion is the marine stratocloud layer (Painemal et al., 2010) that looks prominent when looking down from the peaks of the coastal location observatories. d) The mean elevation and latitudinal extend of the coastal mountain range helps prevent the low-altitude humid air from penetrating inland. The humid air masses from the Pacific are able to penetrate inland only through the occasional valleys by action of thermal winds. However, in the latitude range 19°30' S to 25° S, there is no significant discontinuity in the coastal mountain range (Grenon M., 1990). Certainly, there are factors that contribute to seasonal and annual variability, but in the overall the area is well known for its extreme dryness (Garreaud, R., 2011, Turner et al., 2012, Lakićević et al., 2016), enabling a large fraction of usable nights for up to 89% of the nights in the year (Otarola & Hickson, 2017).

Thus, the AtLAST initiative Site Working Group presents a summary of the atmospheric conditions at various sites in the Atacama desert region in northern Chile that are of potential use for the deployment of a large-aperture single-dish radio telescope such as AtLAST[4]. However, the results from the data analysis reported in here are potentially useful to any other astronomy projects interested in conducting research from any of these geographic locations in northern Chile.

The sites for which data are being summarized in the report are those listed in Table 1, including their coordinates and geographic altitude. Table 1 also includes the period of time for which data is available, the sampling time resolution, and number of days of effective observations. The last column indicates the percentage of days, in the given period, for which observations were effectively performed. For a visual reference, some of the areas of interest (those nearby the Chajnantor plateau) are shown in Figure 1.

Table 1  Sites and Data Provided for this Study

| Site | Latitude Longitude (degrees) | Altitude (m.a.s.l.)[5] | Observation Period | Time Resolution | Number of Days with Observations | % of Data Coverage in Observation Period |
|---|---|---|---|---|---|---|
| Mackenna Mountain Range (Armazones) | -24.5894 -70.1919 | 3000 | 2004/10/28 2008/02/29 | Variable 1-5 min | 1154 out of 1219 total | 94.7% |
| Cerro Tolonchar | -23.9333 -67.9759 | 4480 | 2005/10/13 2008/03/01 | Variable 1-5 min | 726 out of 870 total | 83.0% |
| Chajnantor Plateau (CBI) | -23.0333 -67.7667 | 5080 | 1999/11/06 2008/06/06 | 5 min | 2416 out of 3135 total | 77.0% |
| Chajnantor Plateau (APEX) | -23.0058 -67.7592 | 5150 | 2006/01/01 2018/09/13 | 1 min | 4512 out of 4638 total | 97.0% |
| Cerro Honar | -23.0833 -67.7667 | 5400 | 2000/10/28 2001/11/13 | 10, 20 min | 246 out of 381 total | 65.6% |
| Chajnantor Peak (CCAT-p) | -22.9860 -67.7403 | 5612 | 2007/01/01 2014/12/31 | 15 min | 2800 out of 2921 total | 95.9% |
| Chajnantor Peak (TAO) | -22.9867 -67.7423 | 5640 | 2009/02/21 2012/06/06 | 1 min | 313 out of 1201 total | 26.1% |

---

[4] Initial plans seek for a radio telescope of about 40 m dish diameter.
[5] m.a.s.l. stands for meters above sea level.

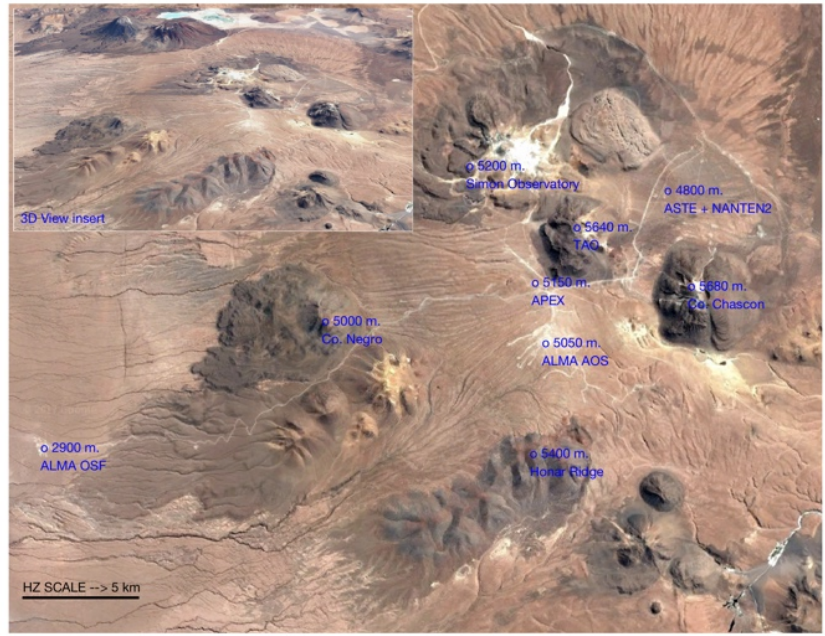

Figure 1 Chajnantor, the location of sites of interest.

## 3. Data: Measurements and Statistical Results

This study focuses only on the analysis of PWV, temperature, wind speed, and wind direction data. These parameters are relevant not only for the design of telescopes and support infrastructure, but also are of interest for the operations of astronomical facilities, as temperature gradients, as well as wind speed and wind fluctuations, may affect the accuracy of pointing of telescopes, especially if not protected by a dome building.

Table 1 lists, for all the sites included in this analysis, the period of time for which data is available. Unfortunately, none of the sites exhibit a 100% of days data coverage. The gaps in the datasets are explained mainly by the remoteness of the sites and poor access roads; occasional snow storms or power outages; or planned maintenance/upgrades of the equipment (e.g. APEX radiometer in 2012), all of which prevent regular access to the sites for the downloading of the data from dataloggers or prevent the data acquisition/data storage process.

The datasets, covering timeseries of atmospheric parameters for the sites in Table 1, are the following. For the Armazones and Tolonchar site, the data are the results of a four years (2004-2008) site testing campaign carried out in the context of site selection for the Thirty Meter Telescope (TMT) project (Schöck et al., 2009). The TMT datasets consist of variable sampling time, but most of the data points correspond to 1-minute averages recorded every 2-minute intervals. The whole TMT site-testing dataset has been made public, and is available through a dedicated website[6] (Otarola et al., 2011a). For the Chajnantor plateau, at the location of the Caltech's Cosmic Background Imager (CBI; Padin et al., 2002) site, the data comes from the weather station installed to support the operations of the telescope. This dataset covers the period from 1999-2007, and consists of instantaneous measurements recorded every five minutes. The dataset is available through the CBI website[7]. The APEX (Güsten et al., 2006) meteorological dataset is available through the project website,[8] and corresponds to data that have been gathered in support of the APEX radio telescope operations. For wind analysis, we have used 1-minute averaged data gathered in the period 2006-2018. For the Chajnantor peak, specifically the location of the Tokyo Atacama Observatory site (Miyata et al., 2008), this study uses a dataset covering the period 2009-2012. Another weather dataset available at a site near the summit of the Chajnantor peak is that gathered in the context of the CCAT (Stacey et al., 2015) and CCATp (Stacey et al., 2018) projects. The dataset[9] corresponding to the period 2007-2014 was made available to this study, and has been previously described in Radford et al., (2008). Finally, wind and temperature data monitored at the Honar site, gathered in the period October 2000-November 2001, was acquired in the context of site studies for the CCAT project in collaboration with the Association of Universities for Research in Astronomy (AURA-Chile).

### 3.1. Precipitable Water Vapor

PWV is undoubtedly a very relevant variable in the determination of sky transmission at various spectral bands of interest for research in astronomy, from microwaves to the mid-infrared. This is particularly true for the AtLAST initiative, which is especially interested in the mm and submm spectral bands.

An analysis of radio soundings launched at 12 UT from the Antofagasta sounding[10] station in the period 2004-2008 was analyzed to get the mean PWV value, and its variability as a function of altitude, the results are shown in Figure 2. Only those soundings with a full record of all the relevant atmospheric variables, up to at least 12 km altitude; were included in the analysis. A total of 762 soundings in the 2004-2008 period met this criteria. As stated in the introduction, the atmospheric conditions in this region of northern Chile are dominated by the subduction of relatively dry air from the southern branch of the Hadley cell. This is a spatially large-scale weather pattern. Consequently, it is fair to assume a relatively low variability in the

---

absolute humidity field in the free atmosphere across the region. This makes the radiosondes dataset launched from the Antofagasta sounding station suitable for this analysis. Also, because of the westerly prevailing wind directions in the upper atmosphere, the radiosondes generally drift in the inland (eastward) direction.

In a previous study, that of Otarola et al. (2011b), the relative humidity (RH) profile of each of the 762 soundings, was corrected to account for a dry bias[11], induced by solar heating of the RH sensor affecting the humidity sensor. The temperature and corrected relative humidity profiles were interpolated to a fix vertical resolution of 25 m, and then converted to a profile of water vapor density following the formalism shown in Section 4 of Otarola et al. (2010). Subsequently, a record of PWV at various sites of interest (Armazones, Tolonchar, Chajnantor Plateau, and Chajnantor Peak) was obtained by vertical integration of each water vapor density profile from the altitude of the sites up to 12 km altitude. The legend included in Figure 2 shows the statistical mean of the 762 PWV values obtained for each site. These results clearly show that at this geographic region, the magnitude of the mean of PWV as well as its variability clearly decrease with elevation, as graphically indicated by the 25%-75% quartiles range in Figure 2.

The insert in Figure 2 shows the statistics of PWV that are known for the Chajnantor plateau site, derived from 225 GHz optical depth measurements conducted during the ALMA site characterization campaign (Radford and Chamberlin, 2000). The procedure consisted of converting the long-term cumulative function of optical depth at 225 GHz[12] using the relationship[13] $PWV [mm] = 23.199*\tau_{225} [nepers] - 0.3142$, that was learned from running the Atmospheric Transmission Model (ATM, Pardo et al., 2001) with profiles of temperature, pressure and absolute humidity that were obtained from radiosonde soundings launched from the Chajnantor site[14].

The PWV derived from the radiosonde data show that the ratio between the mean of PWV for the peak of Cerro Chajnantor and that obtained for the Chajnantor plateau is 0.7, which is in reasonably good agreement with the 0.64 ratio found from a relatively short campaign of simultaneous measurements of PWV at those locations using a 183 GHz radiometer (Bustos, et al., 2014), and matches the 0.7 ratio found from long-term optical depth measurements done with a 350 $\mu$m radiometer (Radford & Peterson 2016). A ratio of 0.64 and 0.7 in the mean PWV measurements between the Chajnantor peak (5612 m altitude location) and the Chajnantor plateau (5080 m altitude location) leads to a water vapor scale height ($h_0$)[15] in the range 1.2 km to 1.5 km. Giovanelli et al. (2001) reports a median water vapor scale height for the Chajnantor site of 1.13 km, which was derived from a series of radiosondes launched during the ALMA site characterization studies. Such scale height corresponds to a PWV ratio of 0.62. However, the water vapor scale height variability is relatively high, ranging from about 0.4 km up to about 2 km (Giovanelli et al. 2001, Figure 2(b)). On the other hand, analysis of daily radiosonde launches from the Antofagasta airport give a mean water vapor scale height (in the section of the troposphere from sea level up to 12 km) of 1.8 km (see best-fit line and legend in Figure 2, or section 2 in Otarola et al., 2011c).

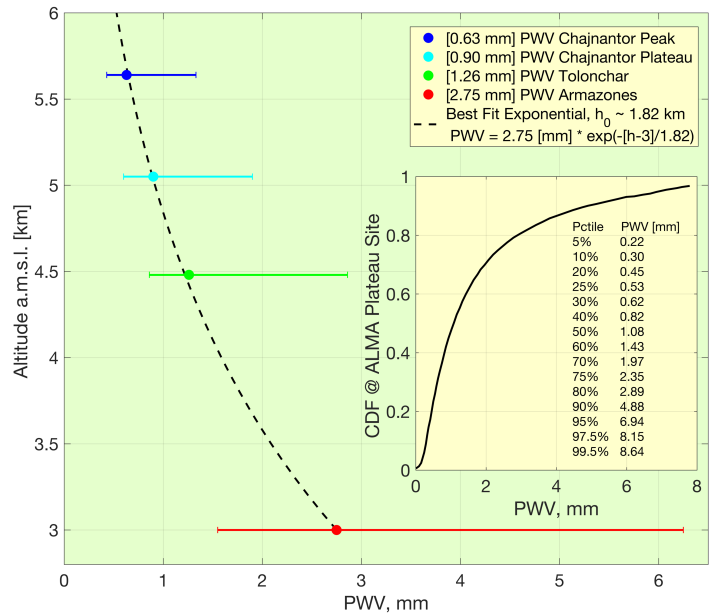

Figure 2 PWV as a function of altitude from 762 radiosondes. The insert shows the long-term statistics of PWV measured at the Chajnantor Plateau. The error bars show the 25 and 75 percentile levels.

The monthly variability of the PWV statistical quartiles is illustrated in Figure 3, and was generated from the analysis of PWV timeseries, covering the period 2006-2017 with 1-minute time resolution, obtained with a 183 GHz radiometer that supports the operations of the Atacama Pathfinder Experiment (APEX) radio telescope[16]. The median value, as shown in Figure 3, seems reasonably stable in the period from March through December covering the fall, winter and spring seasons[17]. Winter storms occurr in the period from May to July, which originate in the development of deep stratiform clouds and precipitation along cold fronts, rooted in low-pressure centers and arching into subtropical laltitudes (Garreaud, R., 2011).

---

[16] See http://archive.eso.org/wdb/wdb/eso/meteo_apex/form
[17] The seasons are defined as for the southern hemisphere.

In austral summer, changes in the regional atmospheric circulation lead to the establishment of the so-called Bolivian high (an upper-level anticyclone), allowing penetration of moist air from the east into the west slope of the Andes region (Garreaud, 2011). This explains the increase in the PWV levels, but just as importantly, implies that this shift in circulation is responsible of stormy weather conditions during the summer and makes the operations of astronomical facilities at the Chajnantor area more difficult. Conditions are generally better on the coastal mountain range, and astronomical facilities such as the VLT keep operating as long as cloudiness is not a problem. The additional cloudiness (about 25%) at the Chajnantor site, due to the advection of moisture in the austral summer, and the increase of cloudiness (about 15%) in the austral winter period is evident in the decrease of clear/usable nights as shown in Figure 3 of Otarola & Hickson (2017).

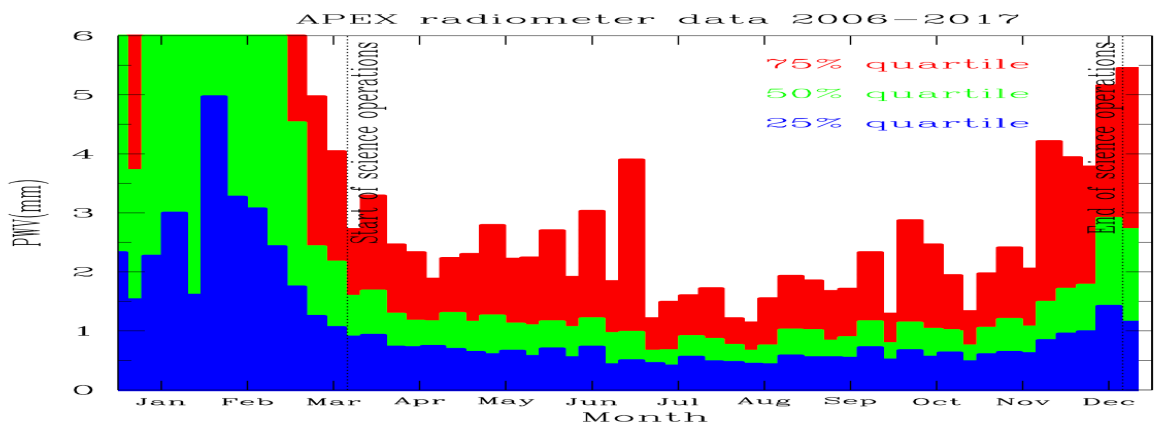

Figure 3 PWV statistics as a function of month at the Chajnantor plateau, as measured using a 183 GHz radiometer at the location of the APEX telescope.

To rank the sites listed in Table 1 regarding PWV, we computed the atmospheric transmission using the *am model version 9.2* (Paine, 2017) setup with the median vertical profiles of water vapor density, temperature and pressure, typical for this tropical location. Then, the am model was run such that the integration of the water vapor profile matched the median PWV for each of the sites of interest. The results are shown in Figure 4. For completeness the atmospheric transmission was computed also for the 10 percentile of PWV. One can infer from the atmospheric transmission that for a median PWV, the Chajnantor plateau offers around 30% transmission in the ALMA bands 9 and 10 (those bands[18] at 602 GHz – 720 GHz and 787 GHz- 950 GHz, respectively), while this increases up to about 65%-70% transmission for very dry conditions that are met 10% of the time. For access to the supra-THz bands, these very dry conditions are absolutely required, while at the Chajnantor plateau, the atmospheric transmission is expected to be around 18%-20%. Placing the telescope at the summit of Chajnantor can help increase the atmospheric transmission in those bands up to 35%-38% level. Cerro Honar, while exhibiting a lower atmospheric transmission than the Chajnantor peak at some spectral bands, offers the advantage of good accessibility and its altitude is below 5500 m. For completeness, for those research teams potentially interested in deploying astronomical facilities, at high geographic altitude in this area, it is worth mentioning to keep in mind that the Chilean law[19] imposes additional requirements, without preventing operations, for projects intending to be deployed at sites above 5500[20] m.

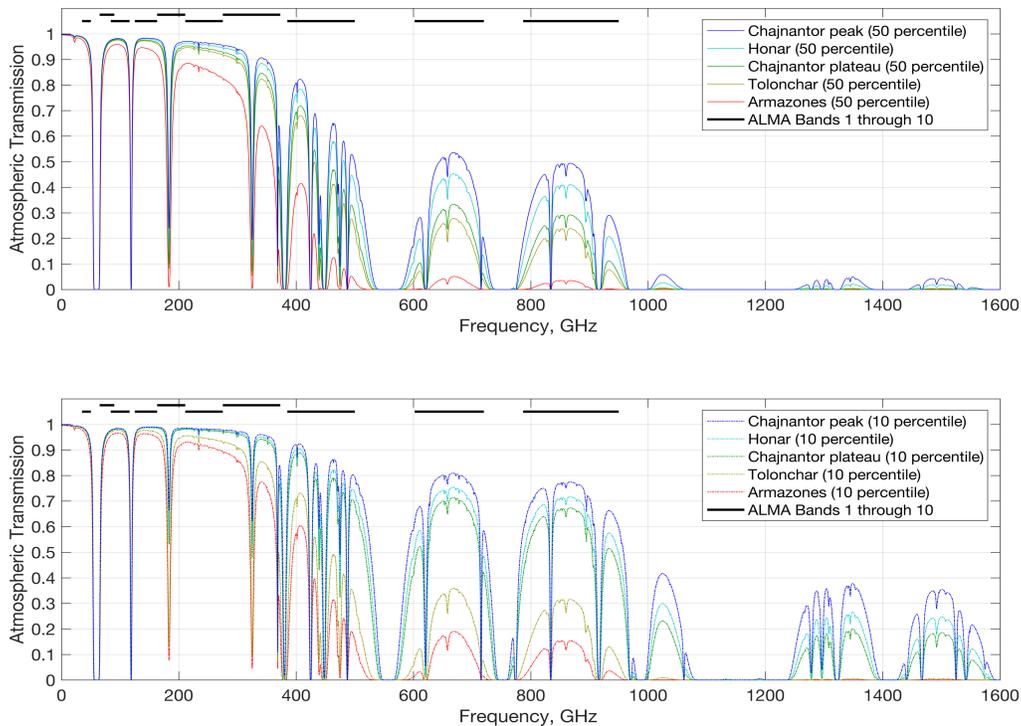

Figure 4 Atmospheric transmission for median (top) and 10 percentile (bottom) PWV conditions at the sites listed in Table 1.

Regarding the daily fluctuations of PWV at the Chajnantor area, the altitude of the atmospheric temperature inversion closer to the ground is very important. During the day, the Sun heats the ground, triggering convection and sublimating water ice from the surface soil. In combination, this increases moisture and helps bring this moisture to higher levels into the atmosphere. In turn, this increases the atmospheric extinction affecting primarily the high frequency spectral bands. At sunset, the lack of Sun radiation stops the convection mechanism and, as the night progresses, colder air subsides bringing the moist air to lower altitudes and increasing the likelihood of relative dry air conditions at the peaks surrounding the Chajnantor plateau. Likely, the best conditions are found during nighttime and winter (Radford & Peterson 2016). This shift in the location of the temperature inversion layers and its effect on the absolute humidity profile is well illustrated in Figure 5. It shows the vertical profiles of temperature and water vapor density for two radiosonde launches, one done in the early morning (red) and one done at night (blue). The radiosondes were launched from the location of San Pedro de Atacama during the ALMA site testing campaign. Other radiosondes were launched from the Chajnantor plateau and the location and variation of the altitude of the temperature inversion has been reported in Giovanelli et al. (2001) and Radford & Peterson (2016). Those studies helped concluding that the Chajnantor peak spend a larger fraction of the time above the near-ground atmospheric temperature inversion layer. The additional contribution of the soundings shown in this study (Figure 5) is that they clearly show that overnight the temperature inversion subsides to altitudes below the plateau bringing the water vapor down with it.

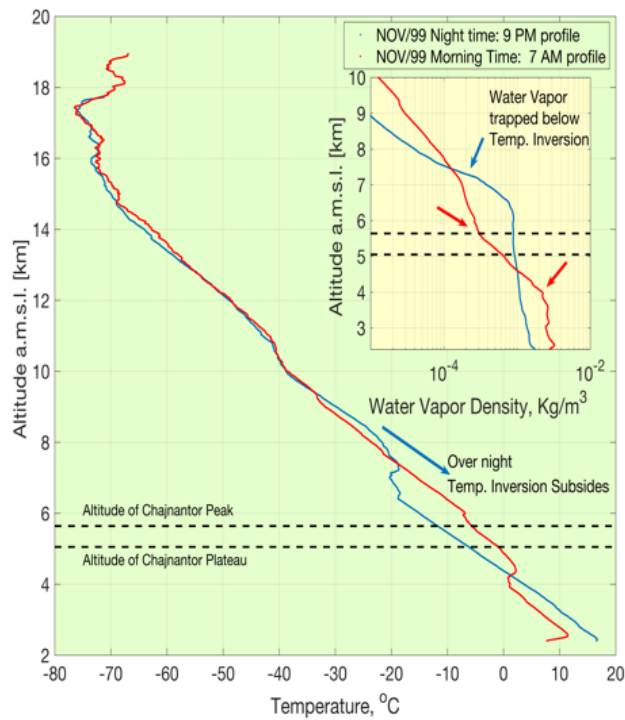

Figure 5 Temperature and water vapor density profiles. Two radiosondes launched (early morning in red, and nighttime in blue) from San Pedro de Atacama.

Temporal variability in PWV on even shorter timescales is relevant to the anomalous refraction effect; especially, for large-aperture radio telescopes[21]. To understand the temporal variability we have used 254 days of PWV measurements, with 5 seconds time resolution, obtained at Paranal with the Low Humidity and Temperature Profiling microwave radiometer (LHATPRO; Kerber et al., 2012).   Out of all the sites in this area, these measurements of PWV are the ones with the highest time resolution, and make them suitable for understanding the PWV variability in short time-scales.

The procedure for the determination of the PWV variability was the following. Each daily timeseries of PWV was split in periods of one hour. Because the time resolution of the observations is 5 seconds, there are a total of 720 measurements per 1-hour record. For each 1-hour record, the standard deviation of the PWV differences (between consecutive measurements), as well as the mean of the PWV, were computed. The PWV fractional variability was computed as the standard deviation of the PWV differences over the mean value. This procedure was repeated for PWV timeseries produced by selecting measurements separated 15, 30, 60, 90, and 120 seconds in time. In a final step, the statistics of the fractional variability, obtained for each hour of the day, and for all the 254 days for which PWV is available, were computed. The statistics include the minimum, maximum as well as the 25%, 50%, and 75% statistical quartiles. The results are shown in Figure 6. In all cases, the results show the daily cycle clearly, where solar heating induced turbulent mixing during daylight hours. The PWV variability reaches its maximum during local midday hours. Other factors, explained earlier, such as variability in the altitude of the atmospheric temperature inversion layer, and sublimation of water ice from the surface soil can also contribute to this daily cycle. This variability increases as the time between measurements increases. The choice of time intervals between 5 and 120 seconds matches the range of usual integration times when observing astronomical sources.

The interquartile PWV fluctuations, as shown in Figure 6, go from about 0.4% to about 1.2% of the mean during 5 s time scales, and this increases up to 1% - 5.5% of the mean PWV for 120 s time scales. If the water vapor field drifts across a 40 m diameter single-dish radio telescope with a mean wind of 5 m/s, it will take 8 seconds for the wind flow to cross the telescope aperture. This information  can prove useful for a subsequent study dedicated  to understand the effect of anomalous refraction in the pointing accuracy of radio telescopes.

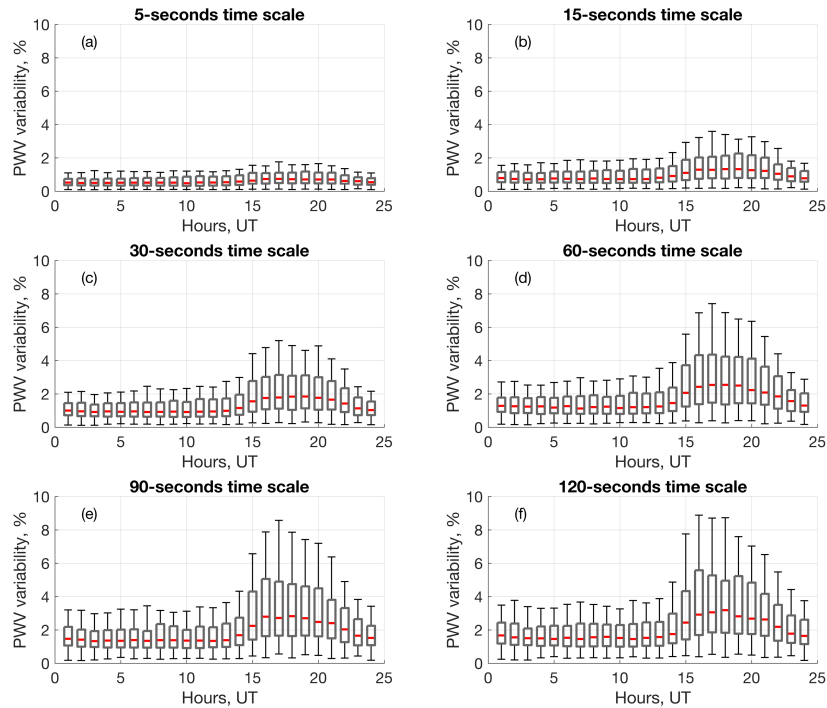

Figure 6 PWV fractional variability for measurements at varying time interval, derived from LHATPRO PWV data at Paranal. Panels (a) through (f), for measurements separated 5, 15, 30, 60, 90, and 120 seconds in time. The vertical scale is the same for all figures with a maximum of 10% fractional variability. The vertical bars show the minimum-maximum range, the rectangular box shows the interquartile (25%-75%) range, and the horizontal line (in red) shows the median value.

## 3.2. Wind Speed and Wind Direction

The wind magnitude has an important effect on the ability of telescopes to achieve a desired pointing stability and accurate tracking of the cosmic sources during the observations and in between pointing calibrations. Consequently, antenna designers pay special attention to this variable. As an example, the requirements set for the design of the ALMA antennas ask for the antennas to meet their performance for a full thermal load and wind magnitude of up to 6 m/s for daytime operations. For nighttime conditions, in the absence of the thermal load from the Sun illumination, the requirements asked for the antennas to meet all performance requirements for wind magnitude up to 9 m/s, with pointing checks every 10 minutes[22]. The ability of the ALMA antennas to meet all performance requirements was verified with the help of Finite Element Modeling (FEM) and throughout testing of two prototypes of the ALMA antennas (Mangum et al., 2006).

---

### 3.2.1. Wind Cumulative Density Function

Figure 7 shows the cumulative density function (CDF) of wind speed, that has been computed from hourly averages of wind speed. Panel (a) includes the statistics for all the sites with the longest data acquisition. On the other hand, panels (b) and (c) show the CDF for the Honar and TAO sites compared to the CDF obtained from contemporaneous meteorological observations acquired at the CBI and CCAT sites, respectively.

In the case of the Honar and TAO sites the number of days with available measurements amount to less than a year, consequently their CDF likely does not capture the multi-year variability typical of atmospheric parameters. However, there is enough overlap between the Honar and CBI sites, as well as between the TAO and CCAT sites, as to get some insights into the spatial variability of the wind between those sites, respectively.

The overlap between the Honar and CBI data acquisition amounts to 246 days, The Honar site, close to the peak of the Honar mountain range, appears windier, with median value about 18% higher, than the CBI located about 300 m below Honar by the Chajnantor plateau. Honar also appears to be specially windier at relatively low wind speeds, this can be inferred from the larger gap between the corresponding cumulative density functions at the lower end of the CDFs.

As for the case of the TAO and the CCAT sites, there are 207 days of contemporaneous meteorological observations, including wind. The TAO site is located on the actual summit of the Chajnantor peak and is open to the prevailing west–north–west winds typical of this area (as shown in Figure 10). The CCAT site is located about 28 m below and about 250 m east from the Chajnantor peak. This difference in altitude, and location toward the lee side of the Chajnantor, help explains the lower wind magnitude statistics shown on Figure 7(a) and 7(c). The statistics obtained from the overlapped and contemporaneous wind data, give a median value of the wind observed at TAO of 5.3 m/s, while at the CCAT site the median value is 3.4 m/s. Overall, the TAO site wind speeds are about 54% larger than those obtained for the CCAT site, but interestingly enough, the TAO site median wind speed is smaller than that obtained for the Honar site. Consequently, our recommendation, for any project interested in deploying infrastructure at the Honar site, is to start by installing a weather station with the goal to monitor, in particular, wind speed for a longer period of time.

A characteristic feature of the TAO cumulative density function is that appears steeper than that of the other sites. In other words, the fraction of the time of very low wind is smaller and the fraction of the time with winds in the 2.5 m/s to 5 m/s is larger than at the other sites. As TAO is at the peak of Chajnantor and relatively open to the winds from the west, this effect shown in the statistics, may be due to compression of the wind flow which is forced by the orography to go over the peak. In other words a manifestation of momentum conservation of the air mass flow.

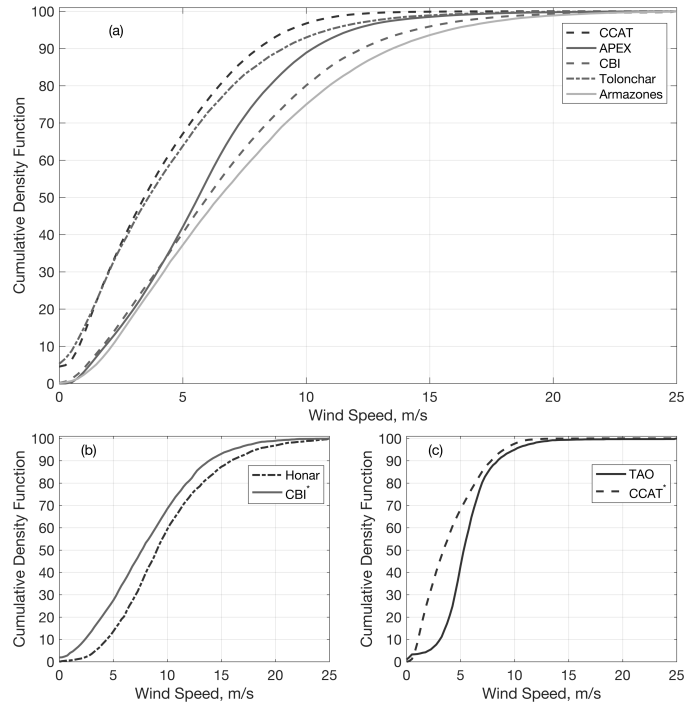

Figure 7 Wind speed cumulative density function (CDF) obtained for each site from 1-hour average wind speed data. Panel (a) shows the CDF for all those sites with the larger record. Panel (b) shows the CDF for the Honar and CBI contemporaneous wind speed data only (about 246 days of data). Panel (c) shows the CDF for the TAO and CCAT contemporaneous wind speed data only (about 207 days of data).

### 3.2.2. Wind Speed Interannual Variability

Asking how long a site testing campaign has to be, as to make conclusive statements about the overall mean conditions of the sites, is a question hard to answer. By definition, for the determination of the climatological mean of a given atmospheric parameter requires 30 years of data[23]. Site testing campaign, with the goal to identify sites of potential use for the deployment of astronomical projects last only a few

---

years. For instance, the site studies intended to rank some sites for the deployment of the Thirty Meter Telescope, lasted four years (between 2004 and 2008).

In this study, as to have an estimation of the interannual variability of the wind speed statistics shown in Figure 7, we have compared some wind speed statistical quartiles of interest, computed using the annual timeseries for the CBI and APEX sites as shown in Tables 2 and 3, against the statistical quartiles computed for the whole dataset (i.e. the long-term statistics) for each site, respectively.

Table 2 CBI site: Overall and Annual Statistics for Assessment of Interannual Variability of the Wind Speed (m/s). GMOD Shows the Geometric Mean of the Differences in the Annual Statistics with Respect to those Obtained for the Whole Data Set (in Percentage).

| Statistics | Data Set | 2000 | 2001 | 2002 | 2003 | 2004 | 2005 | 2006 | 2007 |
|---|---|---|---|---|---|---|---|---|---|
| 25% | 3.4 | 3.5 | 3.8 | 3.0 | 2.9 | 3.5 | 3.5 | 4.0 | 4.1 |
| **50%** | **6.1** | **6.2** | **6.5** | **5.7** | **5.4** | **6.0** | **6.0** | **6.7** | **6.8** |
| 75% | 9.3 | 9.5 | 9.8 | 8.7 | 8.5 | 9.3 | 9.3 | 10.0 | 10.4 |
| 95% | 14.8 | 15.8 | 15.3 | 13.1 | 13.6 | 14.3 | 14.3 | 15.0 | 16.2 |
| mean | 6.8 | 7.0 | 7.1 | 6.1 | 6.1 | 6.7 | 6.7 | 7.3 | 7.5 |
| Std. dev. | 4.4 | 4.8 | 4.4 | 3.8 | 4.0 | 4.1 | 4.1 | 4.4 | 4.6 |
| G.M.O.D | | 3.7% | 2.5% | 9.4% | 10.1% | 2.5% | 2.5% | 4.6% | 10.7% |

Table 3 APEX site: Overall and Annual Statistics for Assessment of Interannual Variability of the Wind Speed (m/s). GMOD Shows the Geometric Mean of the Differences in the Annual statistics with Respect to those Obtained for the Whole Data Set (in Percentage).

| Statistics | Data Set | 2007 | 2008 | 2009 | 2010 | 2011 | 2012 | 2013 | 2014 | 2015 | 2016 | 2017 |
|---|---|---|---|---|---|---|---|---|---|---|---|---|
| 25% | 3.5 | 3.9 | 3.6 | 3.3 | 4.0 | 3.2 | 3.3 | 3.3 | 3.7 | 3.9 | 3.8 | 3.9 |
| **50%** | **5.6** | **6.0** | **5.6** | **5.5** | **6.1** | **5.4** | **5.3** | **5.4** | **5.7** | **5.9** | **5.9** | **5.9** |
| 75% | 7.9 | 8.4 | 7.7 | 2.7 | 8.5 | 7.8 | 7.6 | 7.6 | 8.1 | 8.3 | 8.1 | 8.0 |
| 95% | 12.2 | 12.8 | 11.9 | 12.1 | 13.0 | 12.2 | 11.7 | 11.9 | 12.3 | 12.5 | 12.2 | 12.3 |
| mean | 6.0 | 6.4 | 5.9 | 5.8 | 6.5 | 5.8 | 5.7 | 5.8 | 6.1 | 6.3 | 6.1 | 6.2 |
| Std. dev. | 3.3 | 3.4 | 3.2 | 3.3 | 3.5 | 3.4 | 3.2 | 3.3 | 3.3 | 3.3 | 3.3 | 3.2 |
| G.M.O.D | | 6.1% | 2.4% | 1.6% | 8.0% | 3.0% | 4.1% | 3.1% | 1.8% | 3.8% | 3.5% | 2.9% |

The median wind speed values, from the global and annual datasets, are highlighted. The maximum variability of the annual median wind speed, respect to the whole dataset median, computed for the CBI site is of 11.5%,[24] while the mean of the interannual variability[25] is 6.4%. Similarly, from the APEX wind speed data a corresponding 8.9% and 4.4% has been obtained. For completeness, the geometric mean obtained from the differences (GMOD) between each of the annual statistics shown in Tables 2 and 3, with respect to those obtained from the global datasets are included. Therefore, if the sampling time is high enough, of order of minutes, there will be enough data points collected through a year, such that the overall median value will be of order 5% - 10% with respect to a relatively longer timeseries. This interannual variability may be attributed to year-to-year changes in the main atmospheric circulation patterns over a region, such as El Niño or La Niña oceanic/atmospheric oscillations, or a stronger or weaker Bolivian winter as is the case in this region of the Atacama Desert.

### 3.2.3. Wind Speed: Vertical Gradient in the Near-surface Atmospheric Layer

Also of interest for the design of a telescope is to know the magnitude of the wind speed increases with altitude above ground surface level. Measuring the wind speed gradient requires installation of towers of suitable heights. In the case of the TMT site-testing work, towers of 30 m height above ground level were installed, and wind was measured at near surface (2-m above ground) as well as at 10-m, 20-m, and 30-m above ground. The wind profile, as measured by the TMT site testing campaign, showed an increase of about 8% (median) in the wind magnitude when going from 2-m up to 10-m along the vertical axis, and the median wind speed from 10-m up to 30-m above ground was fairly constant (Schöck et al., 2009).

As for the case of the ALMA site testing, no high tower for permanent monitoring of the wind speed was available. Therefore, in order to have an idea of the wind speed vertical gradient we have analyzed wind profiles obtained from radiosonde soundings. Typically, wind speed decreases near the surface due to loss of energy induced by friction of the wind flow with the rough surface. Yet, at places such as the Chajnantor plateau, surrounded by high peaks, katabatic winds (dense cold air masses flowing down the slope from the nearby peaks, typically after sunset) occasionally create the conditions for a high wind flow closer to the surface.

In the years 2000 and 2002, the ALMA site-testing team launched a total of 122 radiosondes. The section of the wind profile from closer to the ground level up to 150 m above ground was extracted for each sounding. In a subsequent step a linear least-square fit was done to each of the profiles. The slope of the linear fit corresponds to the vertical gradient of wind speed. Of the 122 gradients, 90 (~73.8%) give a positive wind speed vertical gradient (see Figure 8). Seven of the soundings give a relatively large magnitude for the wind gradient;, those are shown separately in Figure 8(b). The median vertical wind gradient, out of the 90 cases with positive vertical wind gradient, is 0.10 m/s per 10-m of height. This implies that the wind speed at 50 m above the surface, under median conditions, will be 0.5 m/s stronger than that near the surface. However, for telescope design purposes engineers may be interested in the stronger wind conditions, the results extracted from the ALMA radiosondes gives a maximum vertical wind speed gradient of 0.66 m/s per 10-m of height (i.e. the wind speed is 3.3 m/s larger at 50-m compare to the near-surface wind conditions).

---

[24] This percentage is the result from taking the largest annual median wind speed value and compare that with the median wind speed obtained from the global data set, (6.8 − 6.1)/6.1*100 = 11.5% rounded to one significant digit.

[25] $\Sigma$\{ |(annual median) − (global dataset median) |\} / (global dataset median) *100%, the summation is over all annual medians.

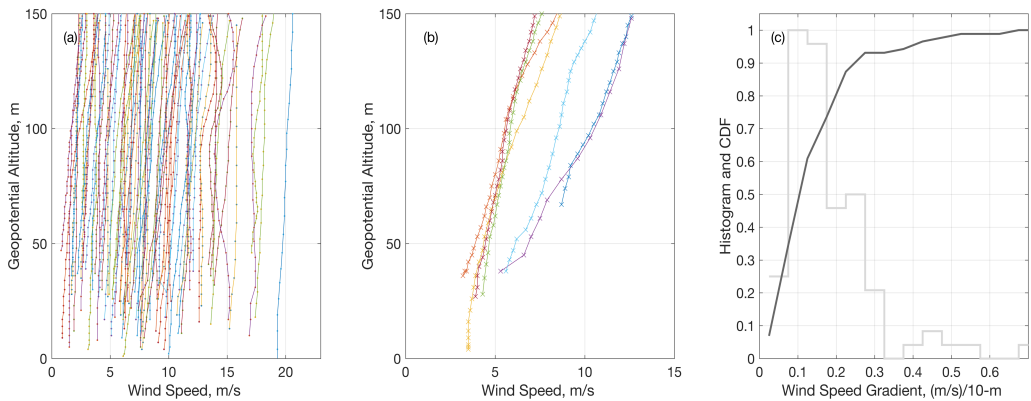

Figure 8 Wind speed as a function of altitude above surface (Chajnantor Plateau) from 122 radiosonde soundings launched in the years 2000 and 2001. Panel (a) shows the soundings (84 in total) in which the wind speed gradient in the section from surface up to 150 m (above ground level) is positive and up to 0.25 (m/s) per each 10 m of altitude (i.e. wind speed increases with altitude). Panel (b) shows only those cases in which the vertical gradient of wind speed (in the section from surface up to 150 m above ground level) was the largest and higher than 0.25 (m/s) per each 10 m of altitude (6 cases). Panel (c) Shows the histogram and cumulative density function (CDF) of the wind speed vertical gradient in units of (m/s) per each 10-m of vertical distance. The median vertical wind gradient is 0.10 m/s per each 10-m of height. The maximum vertical wind speed gradient from the radiosonde soundings give 0.66 m/s per each 10-m of height.

### 3.2.4. Estimation of the Maximum Wind Expected for a 50 Years Recurrence Period

In the design of a telescope, knowledge of the maximum winds the telescope needs to sustain and survive is very important. For this purpose, the CBI wind database, consisting of instantaneous measurements, logged in average at 5-minute interval, is found to be suitable to estimate the wind magnitude that may be expected for period of time comparable to the lifetime of astronomical projects. Instantaneous measurements are able to capture wind gusts and this makes these wind speed measurements specially suitable.

In particular, the analysis done here estimates the magnitude of the wind expected for a 50 years recurrent period. The procedure consisted on extracting, from the CBI wind database, the maximum daily wind for each day for which data is available. In a subsequent step, the daily maximum wind data series was ordered in descending order, from the maximum to the minimum value of the maximum daily wind found in the CBI wind dataset. To each of the measurements we associated a period of recurrence given by $(N_T+1)/n$, where $N_T$ is the total number of data points, and $n$ is the position of a given measurement in the descending order arrangement. This implies the minimum value recorded can happen every day, but the maximum wind speed ever recorded took the number of days, equal to the length of the database in days, to be observed. Figure 9 shows the daily maximum wind speed as a function of recurrence period. Also shown are the marks corresponding to the 50 and 200 years recurrence period, respectively. Extrapolating the results obtained from the CBI data to the 50 years recurrence period mark seems to indicate that a likely daily maximum wind in a 50 years recurrent period is about 51 m/s (this converts to 183.6 km/h, or 114.1 mph[26]). The probability that a 50 years recurrent event will occur in any given year is 1/50*100% = 2%. We may ask what is the probability that such an event will indeed happen in the 50 years lifetime of the project. To answer this question we start by acknowledging that the probability that the event will not happen in a given year is 100%-2% = 98%. Consequently, the probability that such an event won't happen at all in a period of 50 years is 0.98^50*100% = 36.4%. Therefore, the probability that a 51 m/s wind (of 50 year recurrence period) will happen at the site in the lifetime of a project is rather high 100%-36.4% = 63.6%. Perhaps engineers may consider designing for a lower probability event. Our results show that the 200 years recurrence wind speed is 54.3 m/s (195.5 km/h, or 121.5 mph), and there is a 22.2% probability for a 200 years maximum wind speed to happen in the first 50 years of operation of a telescope.

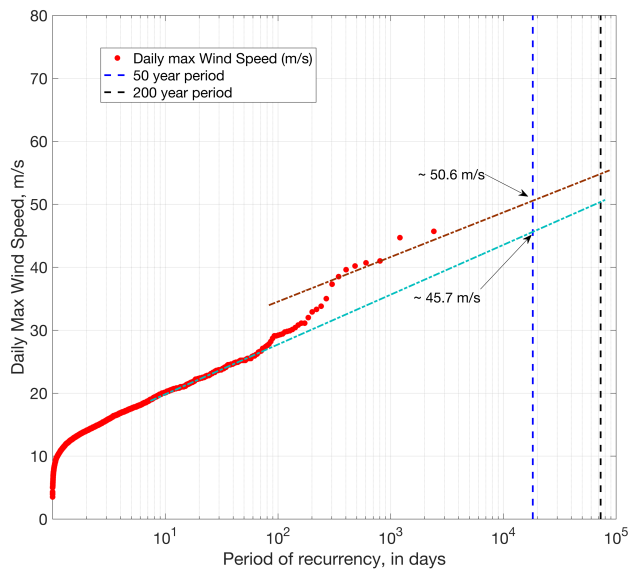

Figure 9  Period of recurrence for daily maximum wind speed at the Chajnantor plateau (CBI site).

[26] mph: miles per hour.

### 3.2.5. Prevailing Wind Direction

Regarding wind direction, Figure 10 shows the windroses for the sites representing the different geographical areas and altitudes in this study. While at the coastal mountain range, and relatively lower elevation sites, the prevailing wind direction at surface level is from the north direction, the higher elevation sites on the west slope of the Andes mountain range shows a WNW[27] prevailing wind direction. Yet, at the location of the APEX telescope the prevailing wind direction exhibits some variability. At nighttime, in the period from about 9 PM to 8 AM local time, the median of the wind direction[28] is 320º (left panel, Figure 11), between 8 AM and 1 PM, the median of wind direction is 282º (center panel in Figure 11), and in the period from 1 PM to 9 PM the median wind direction is 266º and clearly shows a WSW[29] wind direction component (right panel, Figure 11). During the summer period, under the influence of the so-called Bolivian winter, there is a reversal of the wind (a monsoon period in South America), and there is a dominant component from the east, as is possible to see in Figure 10 for the windroses of those sites on the west slope of the Andes mountain. The prevailing wind directions can provide important design considerations when building new telescopes, for example to decide the optimal azimuth for the stow position during strong storms.

## 3.3 Air Temperature

### 3.3.1 Air Temperature Cumulative Density Function

The temperature cumulative density function computed from the temperature record for all sites included in Table 1 is shown in Figure 12. Clearly, the colder site is that located at the highest geographical altitude (the Chajnantor peak), while the warmer conditions are found at the lowest elevation site (Armazones). This is expected from the adiabatic expansion of the atmosphere, where the temperature decreases with altitude. The insert to Figure 12 shows the long-term median temperatures for each site as a function of altitude above sea level. The absolute value of the slope of the linear trend in the median temperature as a function of altitude is the temperature lapse rate and comes to be 6.0 °C/km. In other words, the higher the site is, the atmosphere is not only dryer as shown in Figure 2, but also colder. The temperature record shows a fairly stable variability with a standard deviation of order 4.8 °C ± 0.5 °C at all altitude levels. The interquartile range (75% - 25%) is of smaller magnitude at the Armazones site, 5.3 °C, and of the order of 6.5 °C - 7.5 °C for the high-elevation sites. The median and mean temperatures for the sites at the

---

[27] WNW: West–North–West
[28] Measured clockwise from north.
[29] WSW: West–South–West

Chajnantor plateau and the surrounding peaks are below the freezing point. The relevant temperature statistics are all included in the Table inserted in Figure 12.

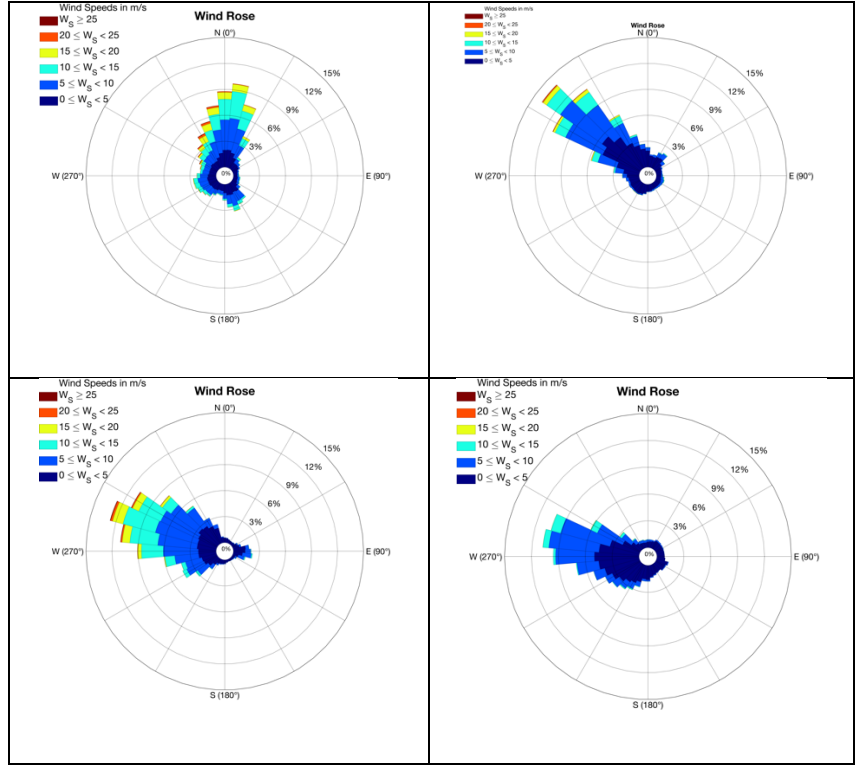

Figure 10 Prevailing wind directions at the Coastal Mountain Range (top left), the Tolonchar site (top right), The Chajnantor plateau (bottom left) and Chajnantor Peak (bottom right).

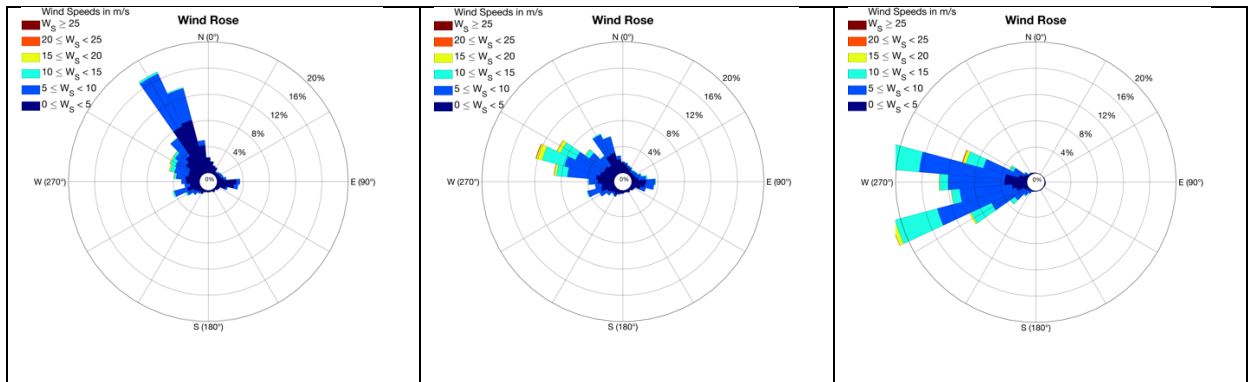

Figure 11 Prevailing wind directions at the APEX telescope site as a function of time of day: (left) nighttime (9 PM to 8 AM local time), (center) morning until early afternoon (8 AM until 1 PM local time), (right) afternoon into early night (1 PM until 9 PM local time)

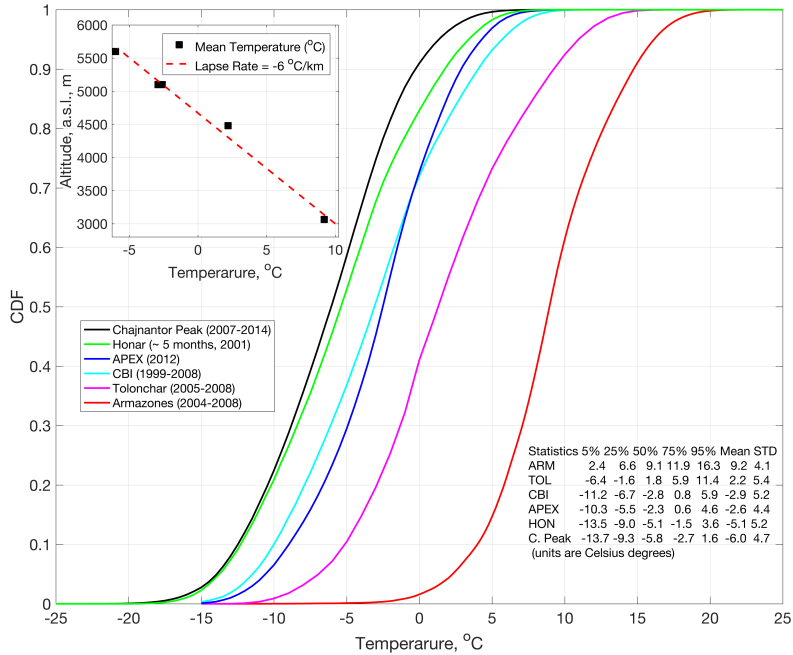

Figure 12 Air Temperature cumulative density function for the sites shown in Table 1 (Armazones (ARM), Tolonchar (TOL), Cosmic Background Imager telescope (CBI), Atacama Pathfinder Experiment telescope (APEX), the Honar peak (HON), and the Chajnantor peak (C. Peak)). The Figure includes the overall statistics (whole day), including the following percentiles: 5%, 25%, 50%, 75%, 95%, as well as mean and standard deviation. The insert shows the mean air temperature as a function of altitude for the ARM, TOL, CBI, APEX and Chajnantor Peak with the segmented line corresponding to a lapse rate of -6 °C/km.

### 3.3.2    Temporal Temperature Gradients

A structure, such as a telescope, exposed to the environmental conditions will be subjected to changes in air temperature. The relevant time scale of the temperature gradients will depend on the thermal inertia of the materials, i.e. their ability to absorb and conduct the heat. The temperature record for the various sites of interest was analyzed to compute the temperature gradients in one hour time scale, the resulting temperature gradient cumulative density functions for all those sites is shown in Figure 13. The median values in the air temperature gradients are found in the range 0.4 °C/hour to 0.7 °C/hour, and the 75% quartile in the range 0.9 °C/hour to 1.5 °C/hour.

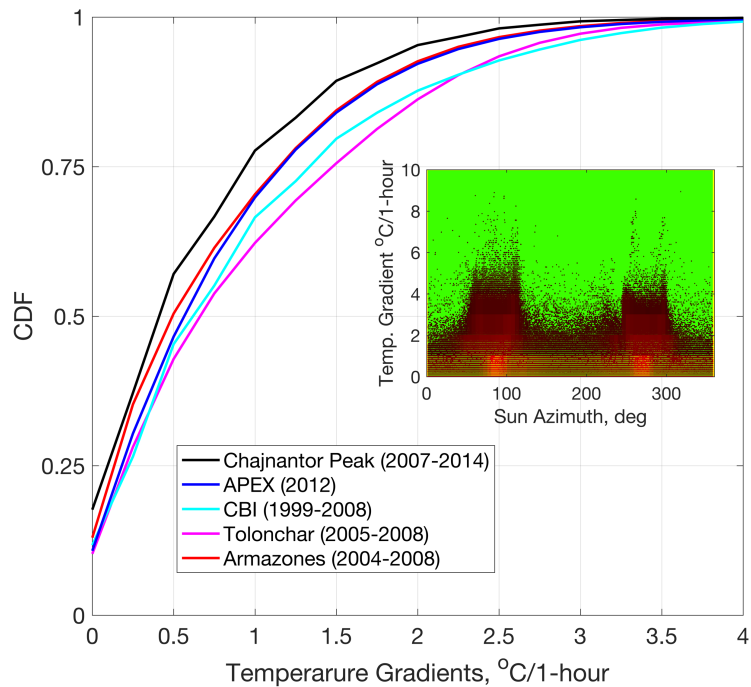

Figure 13 Temperature gradients in 1-hour (sites as shown in Figure 12).

Temperature gradients in a telescope structure, for instance between the left and right supports of the elevation axis could lead to mechanical distortions and fluctuations in the inclination of the elevation axis, introducing changes in the antenna pointing. Antennas, such as the ALMA antennas, are equipped with temperature sensors and highly sensitive inclinometers (Rampini et al., 2010) that are used to monitor the temperature gradients and changes in the inclination of the elevation axis. This information is used to estimate pointing offsets and correct the antenna pointing from those distortions. The APEX telescope team has closely studied the effect of both ambient temperature and temperature near the hexapod (secondary mirror) on the x, y, and z focus of the telescope. They find a clear dependence of the z focus on temperature, which is corrected by a combination of a theoretical prediction augmented by an empirical correction. A weaker dependence on the y-focus (corresponding to the up-down direction) has also been observed, but this has a stronger dependence on elevation than temperature.

The temporal temperature gradients are larger in the transition night/day/night times, and this is confirmed by the data as shown in the insert on Figure 13, which shows the temperature gradients in one hour (blended with the probability map) as a function of the Sun azimuth angle. The temperature gradients are larger when the Sun is rising on the east sky, at azimuth angles between 55º and 120º (in the early morning), and at sunset as shown in the azimuth range 244° - 306° (late afternoon). Therefore, it is advisable to perform pointing and focus calibrations more often at those time of day to reduce the effect of thermal gradients in the pointing and focus of an unprotected single-dish radio telescope.

# 4   Conclusions and Recommendations

Our analysis of the available weather data from the various sites in northern Chile illustrates that there are many parameters to take into account in the selection of the optimal site to install new (sub)mm telescopes, and in particular 40 m-class telescopes without a protecting dome.

The analysis of available data for the sites included in this study, spanning the range of geographical altitudes from 3000 m up to 5600 m above sea level in this region of the Atacama Desert, has been helpful to show that the median value of precipitable water vapor decreases with altitude and so does its variability. Our calculations show that the transparency of the atmosphere  in the microwaves to submillimeter spectral bands is a function of precipitable water vapor. Astronomical projects interested in the spectral bands below 400 GHz may find reasonable conditions at relatively low elevation sites, but if the science goals include observations in the submillimeter spectral bands higher elevation sites, where the median PWV drops to the 1 mm level, will be better. However, projects interested in conducting research with emphasis on the submm spectral bands, specially above 1 THz in frequency, will require of the very dry atmospheric conditions found mainly in the highest peaks in this region. Our conclusion is that projects interested in the submm bands can enjoy of atmospheric transmission of order 35% for at least 10% of the time if located at the Chajnantor peak rather than down at the plateau.

The results shown in Figure 2, for the mean and standard deviation of PWV at the Chajnantor plateau, obtained from  the humidity-calibrated vertical soundings of the atmosphere with radiosondes launched from the Antofagasta station, compare quite well with the long-term statistics known for the site from long-term monitoring of PWV using 183 GHz and 225 GHz radiometers. This seems to imply that on average the water vapor field in the region is quite homogeneous, and the differences between sites are mainly dominated by their geographic altitude. This conclusion may be specially important for the ALMA observatory team, should they consider to enhance the existing array with additional antennas at longer baselines. The results in this study show that sites higher than 4000 m altitude will offer a median PWV below 2 mm (as shown in Figure 2). Therefore, a search for sites at or higher than 4000 m can be done either in a north–south alignment along the west slope of the Andes mountain range, or in the westward direction by looking at convenient locations along the Domeyko mountain range. For instance, Cerro Tolonchar seems to provide adequate conditions for a 100 km baseline in the north–south direction.

PWV variability across the diameter of a radio telescope may be a source of anomalous pointing refraction. High time resolution sampling PWV data show interquartile PWV fluctuations of order 1% of the mean PWV at short time scales (of order 5 seconds), increasing to about 5% of the mean PWV for 120 s time scales. This information has been presented as it may provide support for future studies specially dedicated to understanding anomalous pointing refraction and perhaps phase variability for mm and submm interferometers (such as ALMA).

Regarding wind speed:

1. The Tolonchar and CCATp sites show the best conditions. In the case of the CCATp site, the likely explanation for this lower speed condition is that the site is located on the lee side of the Chajnantor mountain and at a lower altitude than the actual peak. Therefore, this makes the site protected from the prevailing winds. In what the Chajnantor plateau conditions concerns, the wind statistics up to the 50% percentile for the CBI and APEX sites are quite comparable, both located at the Chajnantor plateau. The statistics above the 50% percentile show that the APEX site enjoys of relatively lower speed winds. This may be explained because while the CBI site is open to the environment and rather centrally located in the Chajnantor plateau, on the other hand the APEX site is located towards the northwest edge of the plateau and somehow protected by the Chajnantor mountain range and another small peak (Cerro Chico) located westward from the APEX site. However, the fact that the APEX site is surrounded by higher ground may also explain the variability shown in the prevailing wind direction.

2. Of all the sites studies in this work, the CBI and the APEX sites have the longest data series of wind data. This was helpful to show that the interannual variability of the median wind speed is of order 5%, and the maximum interannual variability was found to be around 10%.

3. Regarding the increase of wind with height above the ground level, radiosondes launched from the Chajnantor plateau show a median vertical wind speed gradient of 0.10 m/s per 10-m of vertical distance, and the maximum vertical gradient learned from the data is 0.66 (m/s) per 10-m of vertical distance.

4. The CBI wind data, consisting of instantaneous measurements of wind speed, and with a record of almost 9 years long, was good to compute the expected maximum speed in a period of 50 years. The likely daily maximum wind to be recorded in a period of 50 years is about 51 m/s (this is 183.6 km/h, or 114.1 mph). The probability that this 50 year recurrent wind will be recorded in a given year is 2%.

The median temperatures calculated for each site gives a temperature lapse rate of 0.6 ᵒC/km. The median values in the air temperature gradients are found in the range 0.4 ᵒC/hour to 0.7 ᵒC/hour, and the 75% quartile in the range 0.9 ºC/hour to 1.5 ºC/hour. The times around sunrise and sunset are the ones exhibiting the stronger temperature radients with temporal temperature gradients reaching 4 ºC/hour are even higher as seen in the inset on Figure 13.

We recommend the installation of a wind station at the Honar site as to verify the windy condition that we have learned from the analysis of contemporaneous wind measurements between the CBI and Honar sites. The current record (only about 240 days of data) may be to short to draw a final conclusion.

Based on the various field campaign in this part of northern Chile, we conclude that PWV, temperature and the general conditions of the wind are well understood. However, when deciding the exact location of a large telescope without protective dome, we advise to study the wind vertical profile in greater detail as well as the wind power spectrum. Such information is relevant to extract requirements for the operational and survival specifications in the structural design.

Based on the atmospheric transparency results, obtained for the median and best 10% quartile of PWV, the best locations for the AtLAST initiative seem to be sites at altitudes of 5000 and above. If the science community behind the AtLAST initiative absolutely requires to maximize the time available for observations in the submm spectral bands, specially above a 1 THz frequency, sites equivalent to the altitude of Honar and Chajnantor Peak are the best options. Both sites, Honar and Chajnantor peaks, are within the boundaries of the Parque Astronómico de Atacama, an area preserved and protected for the deployment of astronomy research projects (Bustos et al., 2014). As for the ALMA observatory, if looking for sites for extending the baselines with the goal to increase the overall effective angular resolution of ALMA observations, then sites at or above 4000 altitude seems suitable, with median PWV at better than 2 mm. The Tolonchar peak offers a good option for a 100 km baseline in the north–south direction.

## *Acknowledgments*

The authors gratefully acknowledge the teams and projects that installed and operated weather stations during site testing and/or operations in the northern part of Chile whose data is either public or has been made available for this article. The projects are the Thirty Meter Telescope (TMT) Project, the Cosmic Background Imager (CBI) of Caltech, the Atacama Pathfinder Explorer (APEX) Telescope, Tokyo Atacama Observatory (TAO), and the C-CAT Project, Atacama Large Millimeter/submillimeter Array (ALMA). In particular, we acknowledge Dr. A.C.S. Readhead, Dr. Kentaro Motohara and Dr. Robert Blum, for providing the meteorological parameters datasets for the CBI, TAO, and Honar sites, respectively. Also importantly, the authors are most thankful to the anonymous reviewer for the throughout review of this manuscript, as well as important suggestion and comments that greatly contributed to improve this manuscript. A. Otarola acknowledges funding from the TMT Project in support of this research.